\documentclass[aps,prl,twocolumn]{revtex4-2}

\usepackage{algorithm}
\usepackage{algpseudocode}
\usepackage{amsmath}
\usepackage{amsthm}
\usepackage{amssymb}
\usepackage{amscd}
\usepackage{bm}
\usepackage{booktabs}
\usepackage{latexsym}
\usepackage{mathtools}
\usepackage{comment}

\usepackage{graphicx}
\usepackage{epsfig}
\usepackage{epstopdf}
\usepackage{svg}
\usepackage[caption=false]{subfig}
\usepackage{placeins}
\usepackage{tabularx}

\usepackage{hyperref}
\usepackage{url}

\renewcommand{\vec}[1]{\boldsymbol{#1}}

\providecommand{\diff}{\mathrm{d}}

\providecommand{\eqref}[1]{(\ref{#1})}

\providecommand{\vanish}[1]{}

\begin{document}
	
\title{A constant of motion for ideal grain growth in three dimensions}

\author{E. Eren}
\affiliation{Department of Materials Science and Engineering, University of California, Davis, CA, 95616, USA.}

\author{J. K. Mason}
\email{jkmason@ucdavis.edu}
\affiliation{Department of Materials Science and Engineering, University of California, Davis, CA, 95616, USA.}

\begin{abstract}
Most metallic and ceramic materials are comprised of a space-filling collection of crystalline grains separated by grain boundaries. While this grain structure has been studied for more than a century, there few rigorous results regarding its global properties available in the literature. We present a new, rigorous result for three-dimensional grain structures that relates the integral of the Gaussian curvature over the grain boundaries to the numbers of grains and quadruple junctions. The result is numerically verified for a grain structure consisting of periodic truncated octahedra.
\end{abstract}


\maketitle

The grain structure of polycrystalline materials is deceptively simple, and for that reason has been the subject of intense and ongoing study. For specificity, consider a model system where the grain boundary energy and mobility are constants, i.e., do not depend on grain misorientation or the boundary plane normal. The phenomenological Turnbull equation \cite{turnbull1951} relates the normal velocity of a grain boundary in such a system to the driving pressure, and along with the Young--Laplace equation \cite{laplace1805} suggests that the normal velocity is directly proportional to the mean curvature of the grain boundary. The migration of individual boundaries induces the evolution of the grain structure, a process known as grain growth, where the total area of grain boundaries and the number of grains decrease with time.

There are surprisingly few rigorous results known about grain structures, even for the two-dimensional version of this system. Energy considerations require that grain boundaries only meet at triple junctions with internal angles of $2 \pi / 3$ \cite{plateau1873,taylor1976}. A consequence of this and curvature-driven grain growth is that a grain's area changes at a rate that depends only on the number of bounding vertices \cite{von1952,mullins1956}. Globally, topological arguments require that the average number of such bounding vertices be precisely six \cite{smith1964}. There are natural analogues to several, but not all, of these results in three dimensions. Grain boundaries only meet at triple junction lines with dihedral angles of $2 \pi / 3$, and triple junction lines only meet at quadruple junction points with angles of $\cos^{-1}(-1/3)$ \cite{plateau1873,taylor1976}. The rate of volume change of a grain depends not only on the total length of the bounding triple lines, but on a measure of the linear dimension known as the mean width \cite{cahn1967,macpherson2007}. For both the two- and three-dimensional systems, the hypothesis that the structure reaches a statistically self-similar state implies that that the average grain diameter increases as the square-root of time \cite{mullins1986,atkinson1988}. This is the effective extent of current knowledge.

There have been a variety of inexact relationships proposed as well, usually for grain structures in the conjectured self-similar state \cite{mason2015}. Ones that relate to the global properties of the three-dimensional system include proposed distributions for the effective radius of a grain \cite{hillert1965,fayad1999,rios2008} and the number of faces bounding a grain \cite{rios2008}. Recent advances in several microscopy techniques promise to make three-dimensional grain structure data more readily available, possibly allowing such relationships to be further refined. Three-dimensional electron backscatter diffraction \cite{uchic2004,konrad2006} destructively images the grain structure by a serial sectioning process, whereas three-dimensional X-ray diffraction microscopy \cite{poulsen2004,li2013} is non-destructive but generally offers poorer spatial resolution. Given this situation, additional rigorous results for the global properties of the grain structure of the model three-dimensional system would be valuable, both to measure deviations of the experimental systems from the model one, and to verify the accuracy of grain structures generated by computational means. This article proves one such result, relating the integral of the Gaussian curvature over the grain boundaries to the numbers of grains and quadruple junction points, and thereby to the numbers of grain boundaries and triple junction lines.

\begin{figure}[b]
	\centering
	\includegraphics[width=0.31\textwidth]{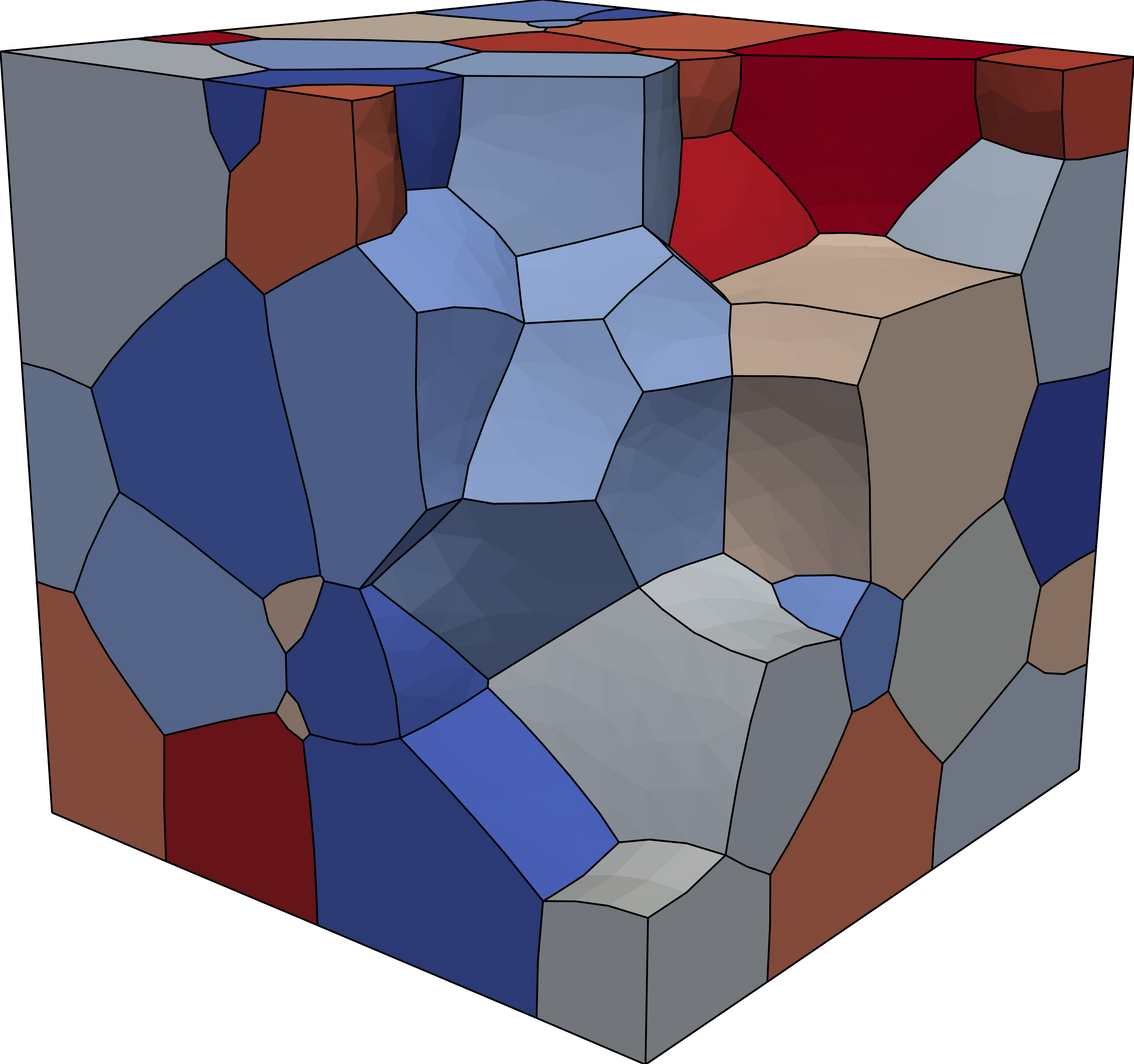}
	\caption{A grain structure in a cubic volume, with several grains removed to reveal the interior. Color indicates the individual grains, internal curved surfaces are grain boundaries, internal black lines are triple junction lines, and four triple junction lines intersect at quadruple junction points.}
	\label{fig:grain_structure}
\end{figure}

Let $\Omega$ be a space-filling grain structure composed of grains that meet in twos on grain boundaries, grain boundaries that meet in threes at triple junction lines, and triple junction lines that meet in fours at quadruple junction points, as in Fig.\ \ref{fig:grain_structure}. Further suppose that $\Omega$ satisfies Plateau's laws (i.e., grain boundaries meet at dihedral angles of $2 \pi / 3$ and triple junction lines meet at angles of $\cos^{-1}(-1/3)$), and that $\Omega$ is defined in a three-dimensional region with periodic boundary conditions. If $G$ is a grain in $\Omega$, then our main result is that the expectation value of the Gaussian curvature $K$ integrated over the interiors of the grain boundaries of $G$ and the expectation value of the number of quadruple junction points $f_{0}(G)$ of $G$ are related by:
\begin{equation}
\bigg\langle \int_{\partial G}{K \: \diff A} \bigg\rangle = 4 \pi - \alpha \langle f_{0}(G) \rangle.
\label{eq:main_result}
\end{equation}
The angle brackets indicate an average performed over all grains in $\Omega$, $\partial G$ indicates the interiors of the grain boundaries of $G$, and $\alpha = 2 \pi - 3 \cos^{-1}(-1/3)$ is the angular defect at a quadruple junction point. This result is exact (given a few technical assumptions that are usually satisfied and are discussed in the supplemental material \cite{CoM_supplemental}), and to our knowledge does not appear in the literature; a related result by Kusner \cite{kusner1992} requires that all the grain boundaries be minimal surfaces, and one by Glicksman \cite{glicksman2005} applies only to unconstructable grain structures of average $n$-polyhedra.

The Gaussian curvature of a surface is defined as the product of the principal curvatures at any point. The appearance of this quantity in Eq.\ \ref{eq:main_result} could be surprising, since the mean curvature (the sum of the principal curvatures) is the one that controls the dynamics of the grain boundary network \cite{laplace1805,turnbull1951}. That said, the Gaussian curvature is in some ways the more fundamental of the two quantities, being an intrinsic property of the surface that does not depend on the way the surface is embedded in Euclidean space. For example, the Gaussian curvature of a sheet of paper is zero at every point whether the sheet is laid flat or rolled up, though the same is not true for the mean curvature. This invariance to the embedding is reflected in the celebrated Gauss--Bonnet theorem:
\begin{equation*}
\int_{\partial G}{K \: \diff A} + \sum_{i = 1}^{f_{2}(G)} \int_{\partial F_i}{\kappa_{g} \: \diff s} + \sum_{i = 1}^{f_{0}(G)}{\alpha_{i}} = 2 \pi \chi(\partial G).
\end{equation*}
While this version specifically applies to the surface of a grain, all versions relate the integrated Gaussian curvature of a surface to its Euler characteristic $\chi(\partial G)$ (equal to two when the surface can be smoothly deformed into a sphere without cutting or gluing). The terms on the left include the integrated Gaussian curvature over the grain boundary interiors, the sum of the integrated geodesic curvature $\kappa_g$ over the interiors of the bounding triple junction lines $\partial F_{i}$ of all grain boundaries $F_{i}$, and the sum of the angular defects $\alpha_{i}$ of the quadruple junction points of $G$.

\begin{figure}
	\centering
	\includegraphics[width=0.22\textwidth]{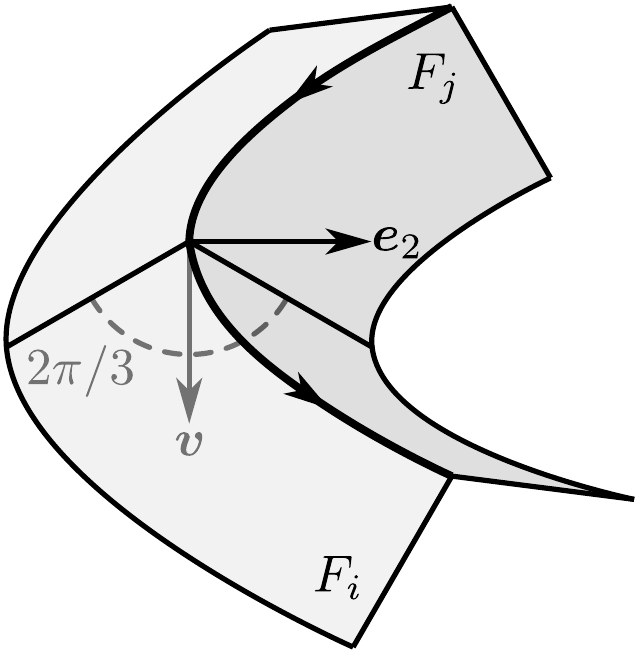}
	\caption{Grain boundaries $F_{i}$ and $F_{j}$ meet at the triple junction line in bold, and $\vec{v}$ bisects the dihedral angle between $F_{i}$ and $F_{j}$. Arrows indicate the tangent direction, and the second Frenet vector $\vec{e}_{2}$ points along the triple junction line's normal direction.}
	\label{fig:tripe_junction}
\end{figure}

If $G$ can be smoothly deformed into a sphere and belongs to a grain structure $\Omega$ that obeys Plateau's rules, then this can be simplified to:
\begin{equation*}
\int_{\partial G}{K \: \diff A} + \sum_{i = 1}^{f_{1}(G)} \int_{E_{i}}{\kappa \vec{e}_{2} \cdot \vec{v} \: \diff s} + \alpha f_0(G) = 4 \pi
\end{equation*}
where the most significant change is to the middle term on the left; this is now the sum of the integrated curvature of the triple junction lines of $G$, weighted by the dot product of the second Frenet vector $\vec{e}_{2}$ of the curve and a unit vector $\vec{v}$ that bisects the dihedral angle between the adjoining grain boundaries; see Fig.\ \ref{fig:tripe_junction}. Summing this equation over all grains in $\Omega$ results in a remarkable cancellation (previously noted by DeHoff \cite{dehoff1994}) where the contribution of the second term on the left vanishes. Specifically, every triple junction line is integrated over three times, once for each adjoining grain. $\kappa \vec{e}_{2}$ is an inherent quantity of the triple junction line that is the same for all three integrals, but the three $\vec{v}$ are all unit vectors in a plane with mutual angles of $2 \pi /3$. That is, the sum of the three $\vec{v}$ vanishes identically for each triple junction line, leaving an alternative version of the main result:
\begin{equation}
\sum_{i = 1}^{f_{2}(\Omega)} \int_{F_{i}}{K \: \diff A} = 2 \pi f_3(\Omega) - 2 \alpha f_0(\Omega)
\label{eq:secondary_result}
\end{equation}
where $F_{i}$ is the $i$th grain boundary of $\Omega$ and $f_{0}(\Omega)$, $f_{2}(\Omega)$ and $f_{3}(\Omega)$ are the numbers of quadruple junction points, grain boundaries, and grains of $\Omega$. Dividing through by $f_{3}(\Omega)$ and multiplying by a constant gives Eq.\ \ref{eq:main_result}. More detailed derivations of both Eqs.\ \ref{eq:main_result} and \ref{eq:secondary_result} are provided in the supplemental material \cite{CoM_supplemental}.

Although Eq.\ \ref{eq:main_result} appears to be simpler, there are at least two observations that are more clearly made by means of Eq.\ \ref{eq:secondary_result}. The first is that the integral of $K$ over the grain boundaries of $\Omega$ depends only on the numbers of grains and quadruple junction points of $\Omega$, and not on the geometry of the grain structure. That is, the left-hand side of Eq.\ \ref{eq:secondary_result} is invariant to any deformation of $\Omega$ that preserves the numbers of grains and quadruple junction points. The second is that a sufficiently accurate measurement of the integral of $K$ over the grain boundaries of $\Omega$ in principle specifies the numbers of all components of $\Omega$. Observe that since there is no rational number that relates the coefficients of $f_{3}(\Omega)$ and $f_{0}(\Omega)$ in Eq.\ \ref{eq:secondary_result}, the numbers of grains and quadruple junctions can be inferred if the left-hand side is known sufficiently accurately. The number of triple junction lines can then be found from $2 f_{1}(\Omega) = 4 f_{0}(\Omega)$ by a counting argument, and the number of grains from $0 = f_{0}(\Omega) - f_{1}(\Omega) + f_{2}(\Omega) - f_{3}(\Omega)$ which follows from the domain of $\Omega$ being a three-torus with $\chi(\Omega) = 0$. The necessary modifications to Eqs.\ \ref{eq:main_result} and \ref{eq:secondary_result} for grain structures in other domains (e.g., ones with free boundaries) are discussed in the supplemental material \cite{CoM_supplemental}.

\begin{figure}
	\centering
	\includegraphics[height=0.17\textwidth]{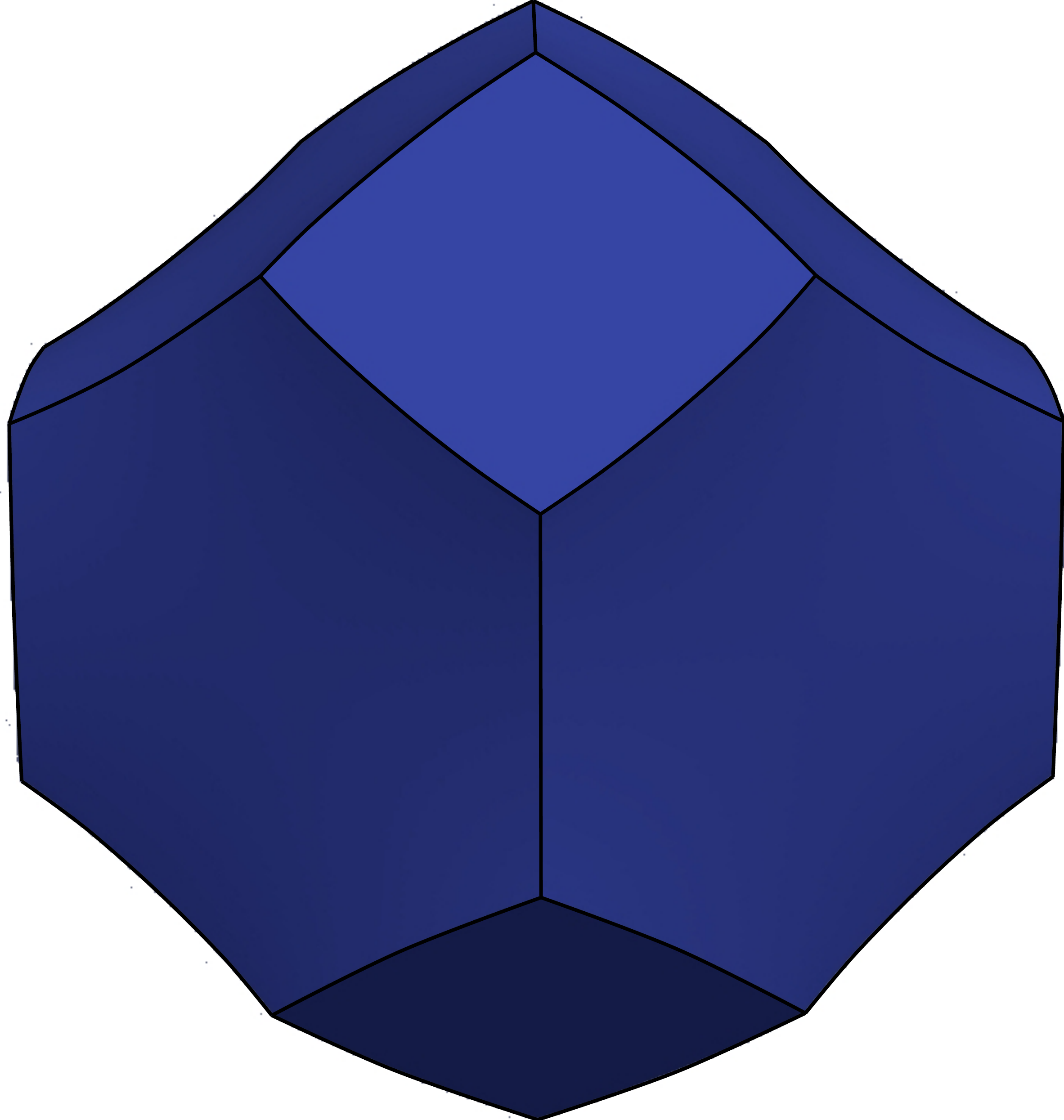}
	\includegraphics[height=0.27\textwidth]{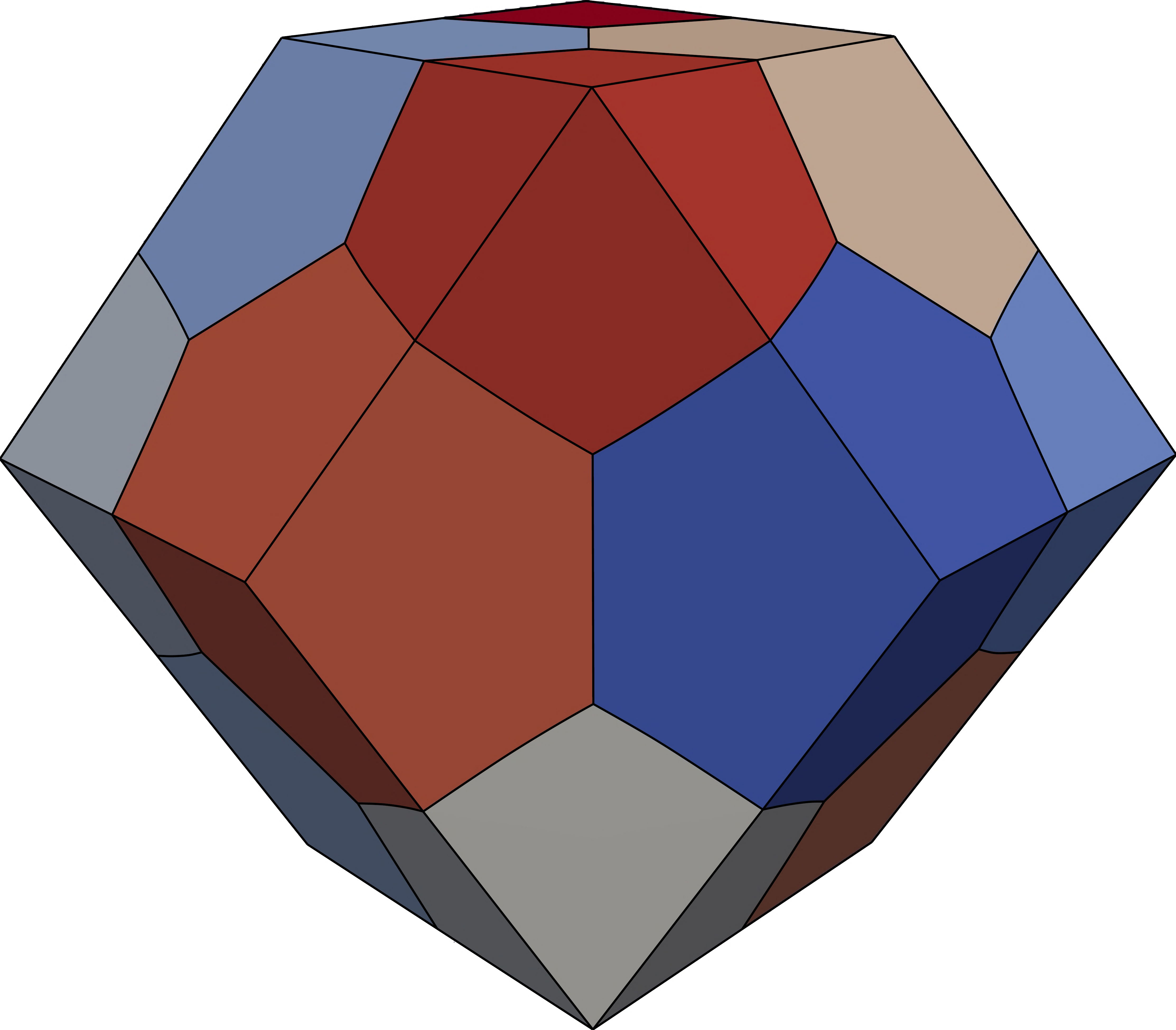}
	\caption{An infinite periodic grain structure that satisfies Plateau's laws can be constructed by repeating a relaxed truncated octahedron (left). This grain was found by starting with a periodic unit of a grain structure consisting of unrelaxed truncated octahedra, fixing the location of the interior quadruple junction points, and minimizing the grain boundary area (right).}
	\label{fig:periodic_structure}
\end{figure}

As numerical confirmation of Eq.\ \ref{eq:main_result}, consider a grain structure consisting of periodic truncated octahedra, relaxed under the action of surface tension to satisfy Plateau's laws; one such grain is shown on the left of Fig.\ \ref{fig:periodic_structure}. The shape of this grain was found by starting with a grain structure consisting of unrelaxed truncated octahedra and identifying the periodic unit shown on the right of Fig.\ \ref{fig:periodic_structure}, with a single grain at the center and corners at the centers of the neighboring grains. The periodic unit was computationally represented by a volumetric finite element mesh with linear elements, with the average number of triangles per hexagonal face $n_t$ depending on a characteristic length and the details of the mesh adaptation algorithm. The locations of the interior quadruple junction points were fixed, and the structure was relaxed by allowing the vertices on grain surfaces to move according to equations of motion known to reproduce curvature-driven grain growth \cite{mason2017} until the magnitude of the vertex forces fell below a threshold. The grain structure did not reach a steady-state configuration when the locations of the interior quadruple junction points were not fixed, owing to a known instability of this grain structure to volumetric perturbations \cite{levine1996}. While quadratic elements would allow the steady-state geometry to be more accurately represented, a convergence analysis with an increasing number of linear elements is sufficient for the present purpose.

The simulations were performed with a modified version of a recently-developed microstructure evolution code \cite{eren2021} that usually uses SCOREC \cite{ibanez2016} for mesh management and maintenance, but the mesh adaptation operations were found to interfere with the convergence of the grain geometry. Instead, artificial vertex forces defined by Kuprat \cite{kuprat2000} were used to maintain the mesh element quality during the structure relaxation. Since the artificial forces only acted on vertices on the grain interiors, it is expected that they did not substantially affect the grain geometry. The boundary conditions were defined to make the simulation cell behave as a periodic unit in a grain structure consisting of periodic truncated octahedra. Whereas the grain boundaries on the simulation cell interior had a constant nonzero energy per unit area, the external surfaces of the simulation cell were assigned zero energy per unit area; this is consistent with viewing them as the result of intersecting grains in the underlying grain structure with the boundary of the periodic unit. Vertices on the external surfaces were constrained to remain on the external surfaces during relaxation by projecting away any displacement in the normal direction, effectively imposing a Neumann boundary condition. The integrated Gaussian curvature was calculated as the sum of the angular defects at the vertices on the grain boundary interiors, where the angular defect is defined as $2 \pi$ minus the sum of the interior angles of the grain boundary triangles meeting at the vertex.

\begin{table*}
    \caption{The percent grain boundary area change of the relaxed truncated octahedron relative to the unrelaxed one and the integrated Gaussian curvature of the relaxed truncated octahedron as functions of $n_t$.} 
    \begin{center}
        \begin{tabular}{@{}ccccccccccccc@{}} \toprule
        $n_t$ & 61 & 129 & 477 & 665 & 1885 & 2610 & 3250 & 4314 & 5322 & 6532 & 7649 & 8906 \\ \midrule
        $\Delta A(\%)$ & 0.0863 & 0.129 & 0.135 & 0.136 & 0.136 & 0.137 & 0.136 & 0.136 & 0.136 & 0.136 & 0.136 & 0.136 \\
        $\int K \: \diff A$ & -0.219 & -0.349 & -0.461 & -0.477 & -0.544 & -0.554 & -0.563 & -0.568 & -0.578 & -0.588 & -0.593 & -0.597 \\
        \bottomrule
        \end{tabular} \label{table:surf_red}
    \end{center} 
\end{table*}

\begin{figure*}
	\centering
	\includegraphics[width=0.70\textwidth]{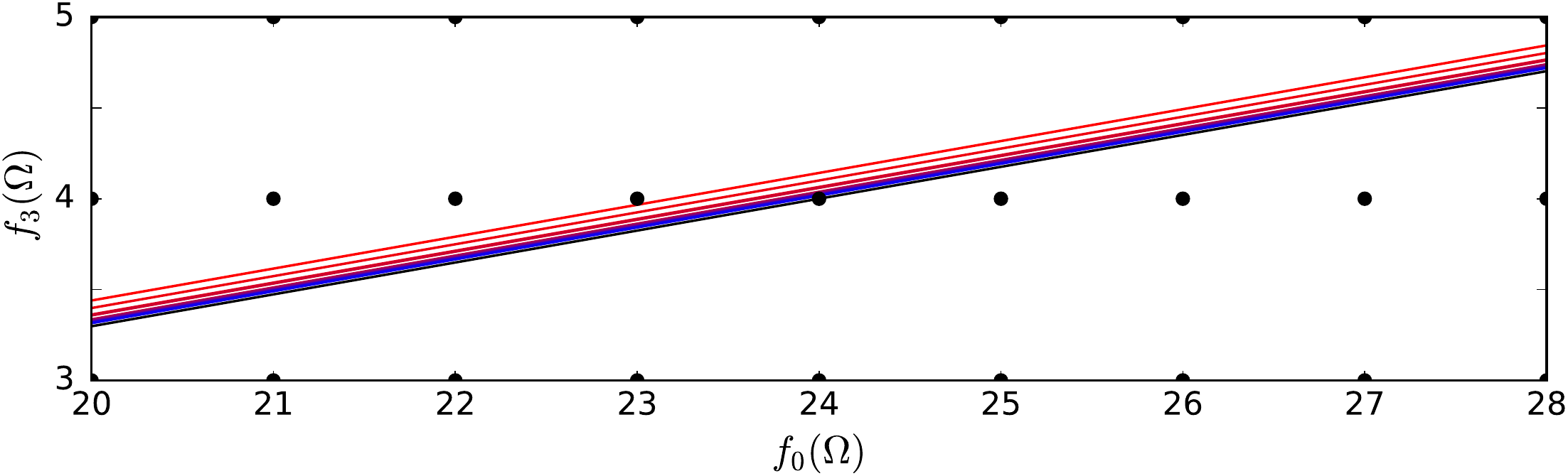}
	\caption{The lines defined by Eq.\ \ref{eq:secondary_result} for the data in Table \ref{table:surf_red} in the feasible region $3 \leq f_{3}(\Omega) \leq 5$ and $20 \leq f_{0}(\Omega) \leq 28$. The lines are colored from red to blue with decreasing error and the black line passing through $(24, 4)$ corresponds to the exact solution. The closest integer lattice point to the line is $(24, 4)$ for $n_t \geq 477$.}
	\label{fig:error_bounds}
\end{figure*}

Table \ref{table:surf_red} shows the results of this analysis for increasing refinement of the mesh, i.e., as a function of $n_t$. The geometric accuracy of the representation can be evaluated by means of the percent reduction in grain boundary area $\Delta A$ of the relaxed truncated octahedron relative to the unrelaxed one. A detailed analysis \cite{reinelt1993} suggests a value of $0.159 \%$ for the continuous system; that $\Delta A$ does not converge to this value is likely due to the irregularity of the mesh. As for the integrated Gaussian curvature, the average quantities in Eq.\ \ref{eq:main_result} are equivalent to those for a single grain by periodicity. This implies that the integral of the Gaussian curvature over the interiors of the grain boundaries should be:
\begin{equation*}
\int_{\partial G}{K \: \diff A} = 4 \pi - 24 \alpha \approx -0.664484.
\end{equation*}
A conjugate gradient minimization algorithm and bootstrapping were used to fit the model $\int K \diff A = a + b n_t^c$ to the data in Table \ref{table:surf_red}, giving $a = -0.661 \pm 0.022$, $b = 2.08 \pm 0.51$, and $c = -0.378 \pm 0.060$ (reported as the medians and half the interquartile range). This implies that the integrated Gaussian curvature would be $-0.661 \pm 0.022$ in the $n_t \rightarrow \infty$ limit, and is interpreted as numerically confirming Eq.\ \ref{eq:main_result} given the degree of approximation of the grain geometry. That the integrated Gaussian curvature converges to the expected value even though the percent area reduction does not confirms the assertion that Eqs.\ \ref{eq:main_result} and \ref{eq:secondary_result} are invariant to geometric perturbations of the structure, provided the numbers of grains and quadruple junction points remain the same and Plateau's laws are satisfied.

Alternatively, one could consider the feasibility of inferring $f_{3}(\Omega)$ and $f_{0}(\Omega)$ by means of a sufficiently accurate measurement of the integral of $K$ over the grain boundaries of $\Omega$ in Eq.\ \ref{eq:secondary_result}. This can be done by a graphical construction in the plane with $f_{3}(\Omega)$ and $f_{0}(\Omega)$ on the vertical and horizontal axes. Given the integral of the Gaussian curvature over the grain boundaries of $\Omega$, Eq.\ \ref{eq:secondary_result} defines a line in this plane with an irrational slope. Since the actual values of $f_{3}(\Omega)$ and $f_{0}(\Omega)$ are necessarily positive integers, this line passes through exactly one point on the integer lattice in the positive quadrant. In practice, any error in the measurement of the integrated Gaussian curvature would change the intercept with the vertical axis and shift the line off of the lattice point; whether this is an issue or not depends on the magnitude of the error and any a priori bounds that can be placed on $f_{3}(\Omega)$ and $f_{0}(\Omega)$. For example, Fig.\ \ref{fig:error_bounds} shows this construction for the data in Table \ref{table:surf_red} with the constraints $3 \leq f_{3}(\Omega) \leq 5$ and $20 \leq f_{0}(\Omega) \leq 28$. Since the magnitude of the error is assumed to be unknown, it is reasonable to suppose that the correct values of $f_{3}(\Omega)$ and $f_{0}(\Omega)$ correspond to the integer lattice point closest to the line within the feasible region. This procedure correctly identifies the relevant integer lattice point as $(24, 4)$ for $n_t \geq 477$; in general, the effect of integrated Gaussian curvature error is reduced as the area of the feasible region decreases.

Apart from advancing our fundamental understanding of grain structures, there remains the question of the practical utility of Eqs.\ \ref{eq:main_result} and \ref{eq:secondary_result} (and the analogues in the supplemental material \cite{CoM_supplemental}). This question is made more pressing by there being few experimental systems that actually evolve by the relevant ideal grain growth process. The authors propose two possible applications based on the differences of the left and right sides of Eqs.\ \ref{eq:main_result} and \ref{eq:secondary_result}:
\begin{align*}
e_1 = \bigg\langle \int_{\partial G}{K \: \diff A} \bigg\rangle - 4 \pi + \alpha \langle f_{0}(G) \rangle \\
e_2 = \sum_{i = 1}^{f_{2}(\Omega)} \int_{F_{i}}{K \: \diff A} - 2 \pi f_3(\Omega) + 2 \alpha f_0(\Omega)
\end{align*}
First, $e_1$ and $e_2$ could be used as rough measures of the deviation of a physical system from an ideal one, along with other quantities like the grain growth exponent and the grain size distribution. Second, $e_1$ and $e_2$ could be used to evaluate the accuracy of the geometric representation of a grain structure in a microstructure evolution code; the derivation above suggests that these quantities should be particularly sensitive to the geometry around triple junction lines and quadruple junction points. Since the angle conditions around triple junction lines are directly implicated in the rates of area and volume change of two-dimensional \cite{von1952,mullins1956} and three-dimensional \cite{cahn1967,macpherson2007} grains, any deviations from Eqs.\ \ref{eq:main_result} and \ref{eq:secondary_result} could function as bounds on the maximum achievable accuracy of simulations of mean-curvature driven grain growth.

\section*{Acknowledgments}
\begin{acknowledgments}
The authors are grateful to R.D.\ MacPherson for enlightening discussions about grain structure geometry. E.E.\ and J.K.M.\ were supported by the National Science Foundation under Grant No.\ 1839370.
\end{acknowledgments}

\section*{Author contributions}
\label{sec:contributions}

J.K.M.\ conceived the study, developed the mathematics, and analyzed the numerical results. E.E.\ designed and implemented the simulation model and performed the numerical experiments. J.K.M.\ and E.E.\ jointly prepared the manuscript.


%

\providecommand{\overbar}[1]{\mkern 3.0mu\overline{\mkern-3.0mu#1\mkern-0.0mu}\mkern 0.0mu}

\onecolumngrid
\newpage
	
\pagebreak
\setcounter{page}{1}
\setcounter{equation}{0}
\setcounter{section}{0}
\setcounter{figure}{0}
\setcounter{table}{0}

\renewcommand{\thetable}{S\arabic{table}}
\renewcommand{\thefigure}{S\arabic{figure}}

\begin{center}
\textbf{\large Supplementary information: \\ A constant of motion for ideal grain growth in three dimensions}
\end{center}

\section{Derivation of Eqs. 1 and 2}
\label{sec:derivations}

Suppose that $\Omega$ is a grain structure in a domain with periodic boundary conditions, that $\Omega$ satisfies Plateau's laws (i.e., grain boundaries meet at dihedral angles of $2 \pi / 3$ and triple junction lines meet at angles of $\cos^{-1}(-1/3)$), and that in $\Omega$ every triple junction line is bounded by two distinct quadruple junction points, every grain boundary is homeomorphic to a disk and bounded by two or more distinct triple junction lines, and every grain is homeomorphic to a ball and bounded by three or more distinct grain boundaries.

The initial objective is to establish the following version of the Gauss--Bonnet theorem for a grain $G$ belonging to $\Omega$:
\begin{equation}
\int_{\partial G}{K \: \diff A} + \sum_{i = 1}^{f_{2}(G)} \int_{\partial F_i}{\kappa_{g} \: \diff s} + \sum_{i = 1}^{f_{0}(G)}{\alpha_{i}} = 2 \pi \chi(\partial G)
\label{eq:gauss_bonnet}
\end{equation}
where $K$ is the Gaussian curvature of a grain boundary, $\partial G$ indicates the grain boundary interiors, $\kappa_g$ is the geodesic curvature of a triple junction line, $\partial F_i$ indicates the interiors of the triple junction lines bounding the $i$th grain boundary, $f_2(G)$ is the number of grain boundaries of $G$, $\alpha_i$ is the angular defect at the $i$th quadruple junction point, $f_0(G)$ is the number of quadruple junction points of $G$, and $\chi(\partial G)$ is the Euler characteristic of the boundary of $G$. The standard version of the Gauss--Bonnet theorem applies to an oriented surface $F_i$ \cite{gray2017}:
\begin{equation}
\int_{F_i}{K \: \diff A} + \int_{\partial F_i}{\kappa_{g} \: \diff s} + \sum_{j = 1}^{f_{0}(F_{i})}{(\pi - \gamma_j)} = 2 \pi \chi(F_i)
\label{eq:standard_gauss_bonnet}
\end{equation}
where $\gamma_j$ is the interior angle at the $j$th corner along $\partial F_i$. As suggested by the notation, let $F_i$ be the $i$th grain boundary of $G$. Since every grain boundary is homeomorphic to a disk, $\chi(F_i) = 1$ and summing Eq.\ \ref{eq:standard_gauss_bonnet} over the grain boundaries of $G$ gives:
\begin{equation*}
\int_{\partial G}{K \: \diff A} + \sum_{i = 1}^{f_{2}(G)} \int_{\partial F_i}{\kappa_{g} \: \diff s} - \sum_{i = 1}^{f_{2}(G)} \sum_{j = 1}^{f_{0}(F_{i})} \gamma_{j} = 2 \pi f_{2}(G) - \pi f_{0,2}(G)
\end{equation*}
where $f_{0,2}(G)$ is the number of distinct pairs of adjacent quadruple junction points and grain boundaries. Adding $2 \pi f_{0}(G)$ to each side of the above equation, using the identity $f_{0,2}(G) = f_{1,2}(G) = 2 f_{1}(G)$, and applying the definition $\chi(\partial G) = f_0(G) - f_1(G) + f_2(G)$ for the Euler characteristic of a surface gives
\begin{equation*}
\int_{\partial G}{K \: \diff A} + \sum_{i = 1}^{f_{2}(G)} \int_{\partial F_i}{\kappa_{g} \: \diff s} + \bigg( 2 \pi f_{0}(G) - \sum_{i = 1}^{f_{2}(G)} \sum_{j = 1}^{f_{0}(F_{i})} \gamma_{j} \bigg) = 2 \pi \chi(\partial G).
\end{equation*}
Since $2 \pi$ minus the internal angles around the $i$th quadruple junction point of $G$ is the angular defect $\alpha_{i}$ at that point, this reduces to Eq.\ \ref{eq:gauss_bonnet} above.

\begin{figure}[b]
\center
\includegraphics[width=0.32\linewidth]{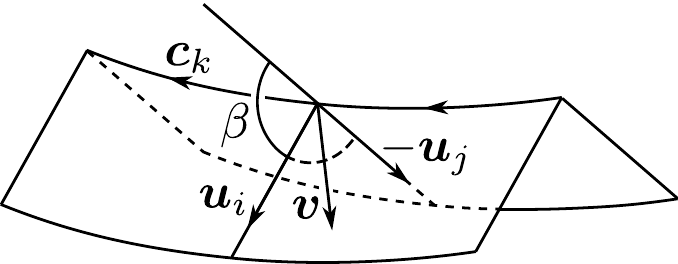}
\caption{The respective arrangement of the curve $\vec{c}_{k}(s)$, the unit vectors $\vec{u}_{i}$, $\vec{u}_{j}$, and $\vec{v}$, and the angle~$\beta$ at a point along $E_k$.}
\label{fig:triple_line}
\end{figure}

The next objective is to simplify Eq.\ \ref{eq:gauss_bonnet} by means of the initial assumptions. Initially observe that $\chi(\partial G) = 2$ since $G$ is homeomorphic to a ball, and that the angular defect at every quadruple junction point is $\alpha = 2 \pi - 3 \cos^{-1}(-1/3)$; this allows Eq.\ \ref{eq:gauss_bonnet} to be reduced to:
\begin{equation*}
\int_{\partial G}{K \: \diff A} + \sum_{i = 1}^{f_{2}(G)} \sum_{j = 1}^{f_{1}(F_i)} \int_{E_j}{\kappa_{g} \: \diff s} + \alpha f_{0}(G) = 4 \pi
\end{equation*}
where $E_j$ is the $j$th triple junction line and $f_1(F_i)$ is the number of triple junction lines of $F_i$. Now consider the summation of the integrals of the geodesic curvature over the triple junction lines of $G$. Every triple junction line is integrated over twice, once for each of the adjoining grain boundaries. Suppose that all of the grain boundaries have outward-pointing unit normal vector fields, that the grain boundaries $F_{i}$ and $F_{j}$ intersect on the $k$th triple junction line $E_k$ of $G$, and that $\vec{c}_{k}(s)$ is the curve parameterized by arc length that travels along $E_k$ in the positive orientation on $F_{i}$ and the negative orientation on $F_{j}$. Let $\vec{\nu}_{i}(s)$ and $\vec{\nu}_{j}(s)$ be the restrictions of the unit normal vector fields of $F_{i}$ and $F_{j}$ to points on $\vec{c}_{k}(s)$, and $\vec{u}_{i}(s) = \vec{\nu}_{i}(s) \times \vec{c}'_{k}(s)$ and $\vec{u}_{j}(s) = \vec{\nu}_{j}(s) \times \vec{c}'_{k}(s)$ be the tangent normals of $\vec{c}_{k}(s)$ on $F_{i}$ and $F_{j}$ as indicated in Fig.\ \ref{fig:triple_line}, where the prime indicates differentiation with respect to arclength. Finally, write the geodesic curvatures $\kappa_{g}$ of $\vec{c}_{k}(s)$ on $F_{i}$ and $F_{j}$ as $\vec{c}''_{k}(s) \cdot \vec{u}_{i}(s)$ and $\vec{c}''_{k}(s) \cdot \vec{u}_{j}(s)$. This allows the middle term on the left of the above equation to be written as:
\begin{align*}
\sum_{i = 1}^{f_{2}(G)} \sum_{j = 1}^{f_{1}(F_i)} \int_{E_j}{\kappa_{g} \: \diff s}
&= 
\sum_{k = 1}^{f_{1}(G)}\bigg[\int_{0}^{s_{k}}{\vec{c}''_{k}(s) \cdot \vec{u}_{i}(s) \: \diff s} - \int_{0}^{s_{k}}{\vec{c}''_{k}(s) \cdot \vec{u}_{j}(s) \: \diff s}\bigg] \\
&= 
\sum_{k = 1}^{f_{1}(G)}\int_{0}^{s_{k}}{\kappa \vec{e}_2 \cdot (\vec{u}_{i}(s) - \vec{u}_{j}(s)) \: \diff s}
\end{align*}
where $s \in [0,s_{k}]$ and $\vec{c}''_{k}(s) = \kappa \vec{e}_2$, or the product of the curvature $\kappa$ and the second Frenet vector $\vec{e}_2$ of $\vec{c}_{k}(s)$. Since $\vec{u}_{i}(s)$ and $\vec{u}_{j}(s)$ are unit vectors, $\vec{u}_{i}(s) - \vec{u}_{j}(s)$ can be written as $2 \sin(\beta / 2) \vec{v}(s)$ where $\vec{u}_{i}(s) \cdot \vec{u}_{j}(s) = \cos\beta$ and $\vec{v}(s)$ is the unit vector pointing along $\vec{u}_{i}(s) - \vec{u}_{j}(s)$ as in Fig.\ \ref{fig:triple_line}. Since $\beta = \pi / 3$, this simplifies further to $\vec{u}_{i}(s) - \vec{u}_{j}(s) = \vec{v}(s)$ and Eq.\ \ref{eq:gauss_bonnet} reduces to:
\begin{equation}
\int_{\partial G}{K \: \diff A} + \sum_{i = 1}^{f_{1}(G)} \int_{E_{i}}{\kappa \vec{e}_{2} \cdot \vec{v} \: \diff s} + \alpha f_0(G) = 4 \pi.
\label{eq:reduced_gauss_bonnet}
\end{equation}
This is perhaps the extent of simplification that is possible for a single grain.

Equations 1 and 2 in the main text are found by summing Eq.\ \ref{eq:reduced_gauss_bonnet} over all the grains in $\Omega$. As described in the main text, the contribution of the second term on the left vanishes since every triple junction line is integrated over three times, once for each adjoining grain. $\kappa \vec{e}_{2}$ is an inherent quantity of the triple junction line that is the same for all three integrals, but the three $\vec{v}$ are all unit vectors in a plane with mutual angles of $2 \pi /3$. This causes the sum of the three $\vec{v}$ to vanish identically for each triple junction line, leaving:
\begin{equation*}
\sum_{i = 1}^{f_3(\Omega)} \int_{\partial G_i}{K \: \diff A} + \sum_{i = 1}^{f_3(\Omega)} \alpha f_0(G_i) = 4 \pi f_3(\Omega)
\end{equation*}
where $f_3(\Omega)$ is the number of grains of $\Omega$. Dividing through by $f_3(\Omega)$ gives Eq.\ 1 of the main text. Alternatively, observing that every grain boundary is included two times in the first sum, that every quadruple junction point is included four times in the second sum, and dividing through by two gives Eq.\ 2 of the main text.

\section{Other boundary conditions}
\label{sec:boundary_conditions}

If the grain structure $\Omega$ exists in a domain without periodic boundary conditions, then the elements of $\Omega$ intersecting the external surfaces need to be handled differently. The derivation of the counterparts of Eqs.\ 1 and 2 of the main text proceeds as above, starting with Eq.\ \ref{eq:gauss_bonnet} for a grain $G$:
\begin{align*}
\sum_{i = 1}^{\overbar{f}_2(G)} \int_{\overbar{F}_i}{K \: \diff A} + \sum_{i = 1}^{\overbar{f}_{2}(G)} \sum_{j = 1}^{\overbar{f}_{1}(\overbar{F}_i)} \int_{\overbar{E}_j}{\kappa_{g} \: \diff s} + \sum_{i = 1}^{f_{2}(G)} \sum_{j = 1}^{\overbar{f}_{1}(F_i)} \int_{\overbar{E}_j}{\kappa_{g} \: \diff s} + \sum_{i = 1}^{\overbar{f}_{0}(G)}{\alpha_{i}} & \\
+ \sum_{i = 1}^{f_2(G)} \int_{F_i}{K \: \diff A} + \sum_{i = 1}^{f_{2}(G)} \sum_{j = 1}^{f_{1}(F_i)} \int_{E_j}{\kappa_{g} \: \diff s} + \sum_{i = 1}^{f_{0}(G)}{\alpha_{i}} & = 2 \pi \chi(\partial G)
\end{align*}
where quantities relating to grain structure elements on external surfaces are indicated by overlines (e.g., the second term on the left concerns triple junction lines on the external surface that bound grain boundaries on the external surface, whereas the third term on the left concerns triple junction lines on the external surface that bound internal grain boundaries). Using that $\chi(\partial G) = 2$ and that the angular defect at every internal quadruple junction point is $\alpha$, repeating the steps involving integrals over triple junction lines, and summing over all the grains of $\Omega$ gives:
\begin{align}
\sum_{i = 1}^{f_3(\Omega)} \sum_{j = 1}^{\overbar{f}_2(G_i)} \int_{\overbar{F}_j}{K \: \diff A} + \sum_{i = 1}^{f_3(\Omega)} \sum_{j = 1}^{\overbar{f}_{1}(G_i)} \int_{\overbar{E}_{j}}{\kappa \vec{e}_{2} \cdot [2 \cos(\lambda / 2) \vec{v}] \: \diff s} + \sum_{i = 1}^{f_3(\Omega)} \sum_{j = 1}^{\overbar{f}_{0}(G_i)}{\alpha_{j}} & \nonumber \\
+ \sum_{i = 1}^{f_3(\Omega)} \sum_{j = 1}^{f_2(G_i)} \int_{F_j}{K \: \diff A} + \alpha \sum_{i = 1}^{f_3(\Omega)} f_{0}(G_i) & = 4 \pi f_3(\Omega)
\label{eq:general_boundary}
\end{align}
where $\cos(\lambda / 2) = \sin(\beta / 2)$ and $\lambda$ is the interior angle along the triple junction line.

Dividing through by $f_{3}(\Omega)$ gives:
\begin{align*}
\bigg \langle \sum_{i = 1}^{\overbar{f}_2(G)} \int_{\overbar{F}_i} {K \: \diff A} + \sum_{i = 1}^{\overbar{f}_{1}(G)} \int_{\overbar{E}_{i}}{\kappa \vec{e}_{2} \cdot [2 \cos(\lambda / 2) \vec{v}] \: \diff s} + \sum_{i = 1}^{\overbar{f}_{0}(G)}{\alpha_{i}} \bigg \rangle & \\
+ \bigg \langle \int_{\partial G}{K \: \diff A} \bigg \rangle + \alpha \langle f_{0}(G) \rangle & = 4 \pi
\end{align*}
which is the counterpart to Eq.\ 1 of the main text and emphasizes the properties of individual grains. This version groups the contributions of grain structure elements on external surfaces in the first average over all grains of $\Omega$. Since $f_{3}(\Omega)$ should increase as the volume whereas the number of terms in the average should increase as the surface area, the influence of the first average should decrease with decreasing surface area to volume ratio of $\Omega$. Alternatively, starting with Eq.\ \ref{eq:general_boundary} and observing that every internal grain boundary is included two times in the fourth term, that every quadruple junction point is included four times in the fifth term, and dividing through by two gives:
\begin{align*}
\frac{1}{2} \bigg \{ \sum_{i = 1}^{\overbar{f}_2(\Omega)} \int_{\overbar{F}_i}{K \: \diff A} + \sum_{i = 1}^{\overbar{f}_1(\Omega)} \int_{\overbar{E}_{i}}{\kappa \vec{e}_{2} \cdot \bigg [ 2 \sum_{j = 1}^{\overbar{f}_{1,3}(\overbar{E}_i)} \cos(\lambda_j / 2) \vec{v}_j \bigg ] \: \diff s} + \sum_{i = 1}^{\overbar{f}_0(\Omega)} \sum_{j = 1}^{\overbar{f}_{0,3}(\overbar{V}_i)}{\alpha_{j}} \bigg \} & \\
+ \sum_{i = 1}^{f_2(\Omega)} \int_{F_i}{K \: \diff A} + 2 \alpha f_{0}(\Omega) & = 2 \pi f_3(\Omega)
\end{align*}
where $\overbar{f}_{1,3}(\overbar{E}_i)$ is the number of grains adjacent to the $i$th triple junction line $\overbar{E}_{i}$ on the external surface and $\overbar{f}_{0,3}(\overbar{V}_i)$ is the number of grains adjacent to the $i$th quadruple junction point $\overbar{V}_i$ on the external surface. This is the counterpart to Eq.\ 2 of the main text. This is as far as the equations can be developed without specifying the shape of the domain of $\Omega$ and the nature of the boundary conditions.

\end{document}